\documentstyle[epsf,rotate,twocolumn,aps]{revtex}

\begin{document}
\draft

\twocolumn[\hsize\textwidth\columnwidth\hsize\csname
@twocolumnfalse\endcsname

\title{
Ideal glass transitions for hard ellipsoids.
}
\author{M. Letz, R.~Schilling, A.~Latz}
\address{Johannes--Gutenberg--Universit\"at, D-55099 Mainz, Germany} 

\date{\today}

\maketitle

\begin{abstract}
For hard ellipsoids of revolution we calculate the phase diagram for
the idealized glass transition. 
Our equations cover the glass physics in the full phase space, for
all packing fractions and all aspect ratios X$_0$. With increasing aspect ratio
we find the idealized glass transition to become primarily be driven by
orientational degrees of freedom. For needle or plate like systems the
transition is strongly influenced by a precursor of a nematic
instability. 
We obtain three types of glass transition lines. 
The first one ($\phi_c^{(B)}$) corresponds to the conventional glass
transition for spherical particles which is driven by the cage
effect. At the second one ($\phi_c^{(B')}$) which occurs for 
rather non-spherical particles
a glass phase is formed which consists of domains. Within
each domain there is a nematic order where the center of mass motion
is quasi--ergodic, whereas the inter--domain orientations build an
orientational glass. The third glass transition line ($\phi_c^{(A)}$)
occurs for nearly spherical ellipsoids where the orientational degrees
of freedom with odd parity, e.g. 180$^o$ flips, freeze independently
from the positions.

\end{abstract}
]
%\vskip -0.5cm
%\twocolumn
\narrowtext
%\begin{multicols}{2}
\section{Introduction}
The dynamics of a molecular system which is supercooled towards the
glass transition shows a variety of phenomena related to the nontrivial
interplay between orientational and translational degrees of freedom
caused, e.g. by steric hindrance.\\
Already in thermodynamic equilibrium molecular systems show, 
compared to simple liquids, a variety
of different physical behavior. At low enough densities (or high
enough temperatures) they form an isotropic liquid. 
On increasing the density 
they can undergo a transition into a crystal 
or several different liquid crystalline phases (like
e.g. a nematic phase). A crucial part of the interaction which causes
these phenomena is given by the shape of the molecules. Therefore one
may also expect new characteristic features for the glass transition
in such systems.\\
A model system which allows to study the translation--orientation 
interplay is a system of 
N {\it hard ellipsoids of revolution} in a box of volume V.
The fluid of ellipsoids is
characterized by two parameters: The aspect ratio
X$_0 = \frac{b}{a}$ relating $b$ and $a$, the major and minor
axis of the ellipsoids, and the packing fraction $\phi$
which is related to the number density
$\rho = \frac{N}{V}$
by $\phi = \frac{1}{6} \pi X_0 \rho$.

In this work we start from a theory of liquids. We demonstrate that
a {\it single} set of equations allows to describe the glassy
behavior for almost spherical particle up to large aspect ratios
where pre-nematic order becomes crucial. 

The choice of the model system has also been motivated by the
successful application of the ideal mode coupling theory (MCT) for
simple liquids \cite{bengtzelius84}, particularly to neutral colloidal
suspensions. 
MCT gives a closed set of equations for the intermediate scattering
function $S({\bf q},t)$.
Comparison between experimental
\cite{vanmegen93} and MCT--results
\cite{bengtzelius84,fuchs92} have shown good agreement for
the colloidal systems which usually are modeled by {\it hard
spheres}. Further tests of the MCT for other
systems can be found in e.g. refs. \cite{goetze91,schilling94,pisa98}. 
MCT in its original form 
describes an idealized glass transition which is indicated by breaking of
ergodicity, at a critical density $\rho_c$ (or critical temperature
$T_c$). The corresponding non-ergodicity parameter $f({\bf q}) =
\lim_{t\rightarrow \infty} {S({\bf q},t)}/{S({\bf q},0)}$ becomes nonzero
at $\rho_c$ (or $T_c$).
Recently the mode coupling equations have been extended to molecular
systems. The dynamics of liquids of rigid molecules composed of M
atoms can either be described by site--site correlators
$S_{\alpha,\beta},\alpha,\beta=1,2,...,M$ or molecular correlation
functions 
$S_{lmn,l'm'n'}({\bf q},t)$
 where for the latter one decomposes
the degrees of freedom into the center of mass and orientational
components (see e.g. \cite{hansen86,gray84}). The density $\rho({\bf
x},\Omega,t)$ is a function of the 
center of mass coordinate ${\bf
x}$ and the orientation $\Omega=(\Phi,\theta,\chi)$ which is specified
by the three Euler angles. 
Expanding  $\rho({\bf x},\Omega,t)$ with respect to a product basis
of plane waves, $e^{i{\bf qx}}$ and generalized spherical harmonics
$D^l_{mn}(\Omega)$ one arrives at the tensorial density
$\rho_{lmn}({\bf q},t)$. $l$ runs over all positive integers including
zero and m and n takes integer values between $-l$ and $l$. Then the
molecular correlators are defined as follows :
\begin{equation}
S_{lmn,l'm'n'}({\bf q},t) = \frac{1}{N} \langle \rho^*_{lmn}({\bf q},t)
\rho_{l'm'n'}({\bf q},0)  \rangle
\end{equation}  
The extension of MCT to molecular systems has been done for the
molecular representation for a single dumb--bell in a simple,
isotropic liquid \cite{franosch97a}, for molecular liquids of linear
molecules \cite{schilling97} and for arbitrarily shaped molecules by
use of nonlinear fluctuating hydrodynamics \cite{kawasaki97} and by
Mori--Zwanzig projection formalism \cite{theis97}. An MCT--approach
using a site--site description has recently been worked out
\cite{sitesite}. Because a hard ellipsoid corresponds to a rigid body
with infinitely many constituents it is the molecular representation
which is the only appropriate one.
Since we consider ellipsoids {\it of revolution} the third
Euler angle $\chi$ becomes redundant. This means that we have to
consider $ S_{lm0,l'm'0}({\bf q};t)$, only. Using the q-frame
\cite{gray84}, i.e. one chooses ${\bf q} = (0,0,q) \equiv {\bf q}_0$
where $q=|{\bf q}|$, these correlators become real and diagonal in m
and m' \cite{schilling97}:
\begin{equation}
S_{lm0,l'm'0}({\bf q},t) = \delta_{m,m'} S_{ll'}(q,m,t) \;\;\;. 
\end{equation}
The head--tail symmetry of
the ellipsoids implies that these correlators vanish for $ l + l'$ odd.
For given X$_0$ the critical packing fraction
$\phi_c$(X$_0$) can be determined by calculating the (unnormalized)
non-ergodicity parameters $
F_{ll'}(q,m) = \lim_{t \rightarrow \infty} S_{ll'}(q,m;t)$.

\section{molecular mode-coupling equations}

Using the densities $\rho_{lm}(q,t)$ and longitudinal translational currents 
$ j^T_{lm}(q,t)$ and rotational currents $ j^R_{lm}(q,t)$ as slow 
variables
set for the Mori-Zwanzig projection operator technique the molecular MCT--equations
have been derived and can be found in refs. \cite{schilling97} and
\cite{theis97}. 
The time dependent molecular MCT--equations can be represented as follows:
\begin{eqnarray}
\label{eq:mctanf}
\frac{\partial}{\partial t} {{\bf S}}(q,m,t) &=&
{{\bf N}}^R(q,m,t)
+ {{\bf N}}^T(q,m,t)
\nonumber \\
\frac{\partial}{\partial t} 
{{\bf N}}^{\alpha}(q,m,t) &=&
 - {{\bf \Omega}}_{\alpha}^2(q,m) \;
{{\bf S}}(q,m,t) \nonumber \\ && \mbox{\hspace*{-1.0cm}}
- \sum_{\alpha'} \nu_{\alpha \alpha'}(q,m) {{\bf N}}^{\alpha'}(q,m,t)
\nonumber \\ && \mbox{\hspace*{-2.0cm}}
- {{\bf \Omega}}_{\alpha}^2(q,m) \sum_{\alpha'} \int_0^t
{{\bf m}}^{\alpha,\alpha'}(q,m,t-t') \;
{{\bf N}}^{\alpha'}(q,m,t) \; dt'
\end{eqnarray}
and:
\begin{eqnarray}
\label{eq:mctend}
m^{\alpha,\alpha'}_{l,l'}(q,m,t) &=& \sum_{{\bf q}_1,{\bf q}_2 \atop
q = |{\bf q}_1+{\bf q}_2|} 
\sum_{m_1} \sum_{l_1,l_2 \atop l_1',l_2'} 
V_{l,l_1,l_2 \atop l',l_1',l_2'}^{m, m_1,\alpha,\alpha'}(q,q_1,q_2) 
\nonumber \\ &&\mbox{\hspace*{0.7cm}}
S_{l_1,l_1'}(q_1,m_1,t) \; S_{l_2,l_2'}(q_2,m_2,t) 
\end{eqnarray}
The indices $\alpha , \alpha' \in \{ T,R \}$ refer to either
translational or orientational currents while $\Theta_{T}$ and
$\Theta_{R}$ are the mass and moment of inertia, respectively. 
${{\bf N}}^{\alpha}(q,m,t)$ are the current--density
correlation functions for translational ($\alpha = T$) and rotational
($\alpha = R$) currents
multiplied with $q$ and $\sqrt{l(l+1)}$, respectively. The
microscopic frequency matrix is denoted by ${{\bf
\Omega}}_{\alpha}(q,m)$ and
is determined by the static molecular correlators. 
In the absence of memory effects (${{\bf m}}^{\alpha,\alpha'}
= {\bf 0}$) the equations are just a set of coupled harmonic
oscillators with friction $\nu_{\alpha \alpha'}$ for vibrational
($\alpha = T$) and
rotational ($\alpha = R$) oscillations. E.g. 
the translational mode with $l=l'=0$ is the propagating phonon mode and the
modes with $l=l'>0$ which exhibit a frequency gap at $q=0$ are {\it localized}
oscillators.

For ${{\bf m}}^{\alpha,\alpha'} \neq {\bf 0}$ nonlinearities
occur. Their physical origin are memory effects.
The corresponding memory kernel is a correlation function of fluctuating
forces. Since fluctuating forces can decay into a {\it pair} of density
excitations this kernel is approximated as a sum of all possible
{\it bilinear} products of density correlation functions. Such a 
non--linear feed 
back mechanism can cause 
an ideal glass transition with non--trivial dynamics.
The glass transition for
eqs. (\ref{eq:mctanf}) and (\ref{eq:mctend}) is investigated in the
following part of the paper.  
The explicit expressions for the vertices $V^{\alpha \alpha'}$ for
arbitrary ${\bf q}$ can be found in ref. \cite{schilling97} and for
the q-frame in ref. \cite{theis97}. 
The vertices $V^{\alpha \alpha'}$ only
depend on the static correlators $S_{ll'}(q,m)$ and the direct
correlation function $ c_{ll'}(q,m)$ which are related to
each other by the Ornstein--Zernike equation.
We have determined $c_{ll'}(q,m)$ within Percus--Yevick approximation.
\\[0.1cm]
It has been shown that the liquid phase of hard ellipsoids is well
described by these approximations \cite{letzallen99}. Although
Percus--Yevick theory fails to describe crystallization it yields a
nematic instability \cite{letzallen99} which is in reasonable agreement with
Monte--Carlo simulations \cite{frenkel84}
even if Percus--Yevick theory still underestimates the tendency
towards orientational order. This nematic instability
will play an important role in the following. For the solution of the
Percus--Yevick equations we have chosen a cut--off $l_{co}=4$ for $l$
and $l'$, whereas the MCT equations were
truncated at $l_{co}=2$. 
We are confident that even such a small number of molecular correlators
enables us to capture the correct physics of the transition.\\[0.1cm]

\section{Solution of the MMCT for hard ellipsoids}

The numerical solution of the eqs. (\ref{eq:mctanf})-(\ref{eq:mctend})
for t$\longrightarrow \infty$ yields 
the non-ergodicity parameters $F_{ll'}(q,m)= {\bf F}(q,m)$. In the limit of
t$\longrightarrow \infty$ the following set of nonlinear equations for
${\bf F}(q,m)$ has to be solved in an iterative way:
\begin{eqnarray}
\lefteqn{
\sum_{\alpha \alpha'} q_l^{\alpha}(q) q_{l'}^{\alpha'}(q)
\left ( {{\bf {\cal F}}(q,m)}^{-1} \right )^{\alpha \alpha'} 
{\bf S}^{-1}(q,m) 
{\bf F}(q,m)} 
\nonumber \\
&&\mbox{\hspace*{1.3cm}}+ {\bf F}(q,m) - {\bf S}(q,m) = 0
\end{eqnarray}
\begin{eqnarray}
{\cal F}^{\alpha,\alpha'}_{l,l'}(q,m) &=& \sum_{{\bf q}_1,{\bf q}_2 \atop
q = |{\bf q}_1+{\bf q}_2|} 
\sum_{m_1} \sum_{l_1,l_2 \atop l_1',l_2'} 
V_{l,l_1,l_2 \atop l',l_1',l_2'}^{m, m_1,\alpha,\alpha'}(q,q_1,q_2) 
\nonumber \\ &&\mbox{\hspace*{1.3cm}}
F_{l_1,l_1'}(q_1,m_1) \; F_{l_2,l_2'}(q_2,m_2) 
\end{eqnarray}
with
\begin{equation}
q_l^{\alpha}(q) = \left \{ 
\begin{array}{ccl}
q & \;\; \mbox{for} & \alpha = T \\  \sqrt{l(l+1)} & \;\; \mbox{for} &
\alpha = R  
\end{array} \right .
\end{equation}
with ${\bf \cal F}(q,m) = lim_{z \rightarrow 0} -z{\bf
m}^{\alpha,\alpha'}(q,m,z) = lim_{t  \rightarrow \infty} {\bf
m}^{\alpha,\alpha'}(q,m,t)$ we denote the long time limit of the
memory kernel. 
From a solution of these equations we obtain
the phase diagram for ideal glass
transitions which is shown in fig. \ref{fig:3}. This figure also
contains two dashed--dotted lines $\Phi_{nem}(X_0)$ indicating the
location of the nematic instability as it arises in thermodynamic
equilibrium from PY-theory \cite{letzallen99}. 
These two lines are in agreement with density functional
theory \cite{groh97} and Monte Carlo simulations \cite{frenkel84}.
In addition there are {\it three} glass transition lines, each for $X_0<1$
and $X_0>1$. First of all we will discuss the critical line
$\phi_c^{(B)}(X_0)$ (thick solid line) at which {\it both}
translational and orientational degrees of freedom for $l$ and $l'$
{\it even} undergo a discontinuous ergodic to non-ergodic transition
(also called type--B transition). The existence of $\phi_c^{(B)}(X_0)$
has been established for $0.35 < X_0 < 2.5$. 
In the region where the $\phi_c^{(B)}(X_0)$
glass transition occurs the equilibrium system
shows crystallization. Being a first order phase transition the onset
of crystallization gives two densities (e.g. from Monte--Carlo
simulations \cite{frenkel84}) resulting from a Maxwell
construction. The $\phi_c^{(B)}(X_0)$ glass transition line is well
bracketed between these two densities.
This indicates that the mode
coupling equations describe a glass transition in the meta-stable
region of a super-cooled liquid. 
The physical origin of the glass
transition depends strongly on the location on
$\phi_c^{(B)}(X_0)$. For aspect ratios $X_0$ close to one the transition is
dominated by the center of mass correlator $S_{00}(q,0)$. 

To
illustrate this point we have plotted in fig. \ref{fig:1} the 
static center of mass correlator and the ''quadrupolar'' correlator 
$S_{22}(q,0)$ and their corresponding non--ergodicity parameters for
X$_0=1.3$.   
This
was done directly above the critical packing fraction
$\phi_c=0.549$.
The first peak at $q_{max}=q\approx 6.6 [a^{-1}]$ of 
$S_{00}(q,0)$ dominates the
transition which is the manifestation of the cage effect. 
Stronger deviations of the ellipsoids from spherical symmetry, however,
alter this behaviour. This is demonstrated in fig. \ref{fig:2} where we
have plotted the same correlators as in fig. \ref{fig:1} but for an
aspect ratio of X$_0=2.3$. Now the peak at $q \approx 0$ of
the quadrupolar correlator $S_{22}^0(q,0)$ (which is for q=0 the Kerr constant
for non polar fluids) dominates the breaking of ergodicity.
The half width $\Delta q$ (at half maximum) of this peak defines a
correlation length 
$\xi = 2 \pi / \Delta q$. In fig. \ref{fig:4} we have plotted $\xi$ at
the glass 
transition line (either $\phi_c^{(B)}$ or $\phi_c^{(B')}$) as a
function of the aspect ratio X$_0$ for prolate ellipsoids.   

Within the glassy phase, i.e. for $\phi > \phi_c^{(B)}(X_0)$ a
continuous (also called type--A) glass transition occurs at the
critical lines $\phi_c^{(A)}(X_0)$ (thin solid lines) at which the
self part of the correlators with $l$ and $l'$ {\it odd}
freezes. This
type--A transition can only occur if the corresponding vertex
$V$ is large enough. For this to happen the aspect ratio should
clearly be different from one. The reader should note that four points
(open circles) were determined exactly and the thin solid line is
schematic, showing that $\phi_c^{(A)}(X_0)$ has to increase if $X_0$
is changed towards one, in order to keep the vertices large
enough. The physical interpretation is that at $\phi_c^{(A)}(X_0)$ the
$180^{\circ}$--jumps of the ellipsoids become frozen. This resembles the
formation of orientational glasses.
One possible candidate for such a transition might be plastic crystals
like the carboranes \cite{brand99} although presently only type B
transitions are detected.

Probably the most interesting result is the third critical line
$\phi_c^{(B')}(X_0)$ (dashed line) which is shown schematically in
figure \ref{fig:3} for $X_0 > 2.0 $ (prolate ellipsoids) and $X_0 <
0.5$ (oblate ellipsoids). In this region the glass transition lines
are close to the nematic instability line. The existence of
$\phi_c^{(B')}(X_0)$ is based on our following
observations. Increasing $\phi$ for $2.1<X_0 <2.5$ we find a glass
transition at $\phi_c^{(B)}(X_0)$ where all $F_{ll'}(q,m)$ become
nonzero. Increasing $\phi$ further we find in addition a second
solution $F'_{ll'}(q,m)$ for $\phi \ge \phi_c^{(B)}(X_0)$. This
solution has the feature that $F'_{ll'}(q,m)$ is practically zero with
the exception of a well pronounced peak for $F'_{22}(q,m)$ at
$q=0$ and with a width of order $\xi$. We have shown this in
fig. \ref{fig:6}a where we have plotted 
$F_{22}'(q,0)$ at the $\phi_c^{(B')}$ transition for $X_0=2.5$. This
is plotted together with the static correlator $S_{22}(q,0)$ and the
normalized function $f_{22}'(q,0) = F_{22}'(q,0) / S_{22}(q,0)$. In
fig. \ref{fig:6}b we have plotted the same quantities for $l=l'=0$,
the center of mass correlator. $f_{00}'(q,0)$ does not exceed $0.15$
although the orientations are frozen.    
This means that the system is ''quasi--ergodic'' in the sense
that for length scales $l \ll \xi$ the ellipsoids show a (nematic)
orientational order and the center of mass behaves ''quasi--ergodic''
decaying to a very small value. Whereas
for length scales $l \gg \xi$ the orientations as well as the
positions have non--decaying, long--time correlations and are frozen. The
easiest way to think of such a system is due to the formation of
liquid crystalline (nematic) domains with a size of the order of
$\xi$. This is visualized in fig. \ref{fig:5}. Fig. \ref{fig:4} shows that the
domains can be quite large. For X$_0 = 2.5$ (the aspect ratio where the
type-B' transition occurs first and where therefore the type-B'
transition with the smallest domain
size shows up) we obtain from our calculation a domain size with $\xi \approx
30$. Within the domains the center of mass is ''quasi--ergodic''
i.e. liquid like 
whereas the orientations are frozen with a nematic order. In our
idealized MCT an ellipsoid can not 
move from one domain to the other. 

In connection with this it is also interesting to mention that 
two types of type B transitions were also
found for the center of mass correlator of a 
{\it simple} liquid of hard spheres
with an attractive interaction given either by the Baxter model
\cite{fabbian98} or a Yukawa potential
\cite{bergenholtz99}. The existence of these two solutions for $X_0 >
2.1$ reflects the competition between  the frozen positional disorder
due to the cage effect  and the tendency to form a nematic
phase. Since $F'_{ll}(q,m) < F_{ll}(q,m)$ for $2.1 < X_0 <
2.5$ the second solution is unphysical \cite{coma}. However, for $X_0
\geq 2.5$ and $\phi_c^{(B')}(X_0) \leq \phi \leq \phi_{nem}(X_0)$ we
only find 
one solution which has all the features of $F'_{ll'}(q,m)$ described
above. 
We stress that the existence of the critical line $\phi_c^{(B')}(X_0)$
trusts on our choice of slow variables which includes the nematic
order parameter and therefore accounts for the occurrence of the weakly
first order nematic transition.
Since $\phi_c^{(B')}(X_0)$
is rather close to $\phi_{nem}(X_0)$
''quasi--critical'' fluctuations appear which also slow down the
entropy fluctuations. We do not think that those will qualitatively
change the phase diagram.
On the other hand the concept of a glass transition
induced by the vicinity of a second order phase transition was already
introduced by a MCT approach \cite{aksienov87} in order to describe
the experimentally observed central peak 
phenomenon close to a ferroelectric instability. 

\section{conclusion}

In conclusion, we have shown that hard ellipsoids exhibit a rather intriguing
phase diagram obtained from the idealized mode coupling theory for
molecular systems, where the orientational degrees of freedom and
their coupling to translational ones is incorporated. Particularly,
we predict a glass transition for $X_0 > 2$ which is driven
by a precursor of a nematic phase. 
Ellipsoids show two type-B glass transition lines
($\phi_c^{(B)}$ and $\phi_c^{(B')}$). One, $\phi_c^{(B)}$,
is dominated by the cage effect whereas the other one $\phi_c^{(B')}$ is caused
by an orientational (nematic) instability.
Besides this a type-A glass transition occurs for almost spherical
ellipsoids where the orientational degrees of freedom with odd parity,
e,g, 180$^o$ -flips freeze independently from the translational ones.
It would be very interesting to
check these predictions by experiments or simulations.  

\acknowledgements
We thank M.~Fuchs and K.~Kawasaki for discussions on the role of the
quasi--critical fluctuations and gratefully
acknowledge
financial support from 
the Deutsche Forschungsgemeinschaft through
SFB 262. 

%\bibliographystyle{unsrt}
%\bibliography{lit_mod}

\begin{thebibliography}{10}
\bibitem{coma}
It was proven in \cite{goetze91} that the largest solution of the non
linear MCT equations is the physical one. It is not clear whether this
can be extended to molecular liquids. However we have found that
$F'_{ll}(q,m) < F_{ll}(q,m) $ for {\it all} $q$,$m$ and $l=l'\leq 2$,
it is tempting to believe that it is still true.
\end{thebibliography}

\begin{figure}
\unitlength1cm
\epsfxsize=10cm
%\begin{picture}(7,7.5)
%\put(-2.6,0){\rotate[r]{\epsffile{fig1.ps}}}
%\end{picture}
\caption{{
Phase diagram for the ideal glass transitions. 
The horizontal axis shows $X_0$ scaled with $\frac{X_0^2-1}{X_0^2+1}$. The 
type-B glass transition lines $\phi^B_c(X_0)$ and $\phi^{B'}_c(X_0)$ 
(see text) are depicted as thick solid and dashed lines, respectively.
The thin solid line is the $\phi_c^{(A)}(X_0)$ glass transition
line. The nematic instability occurs at $\phi_{nem}(X_0)$ which is
shown as thin dashed--dotted lines. The inset shows the situation
around X$_0 = 2.5$ where the $\phi^B_c(X_0)$ glass transition line
merges into the $\phi^{B'}_c(X_0)$ transition line. 
}}
\label{fig:3}
\end{figure}

\begin{figure}
\unitlength1cm
\epsfxsize=10.5cm
%\begin{picture}(7,7.5)
%\put(-2.2,-0.5){\rotate[r]{\epsffile{fig2.ps}}}
%\end{picture}
\caption{{
The static structure factor $S_{ll'}(q,m)$ is plotted
together with the non-ergodicity parameter $F_{ll'}(q,m)$ for 
X$_0$ = 1.3 and $\phi$ = 0.549 (directly above the non-ergodicity
transition). Part a)
shows the center of mass correlator $l=l'=m=0$ whereas part b) shows
the quadrupolar correlator $l=l'=2, m=0$.
}}
\label{fig:1}
\end{figure}

\begin{figure}
\unitlength1cm
\epsfxsize=10.5cm
%\begin{picture}(7,7.5)
%\put(-2.2,-0.5){\rotate[r]{\epsffile{fig3.ps}}}
%\end{picture}
\caption{{
A similar plot as fig. \protect\ref{fig:1} is shown but for X$_0$ = 2.3
and $\phi = 0.617$ (again directly above the non-ergodicity
transition)
which is already close to the nematic instability.
}}
\label{fig:2}
\end{figure}

\begin{figure}
\unitlength1cm
\epsfxsize=10.5cm
%\begin{picture}(7,7.5)
%\put(-2.2,-0.5){\rotate[r]{\epsffile{fig3.ps}}}
%\end{picture}
\caption{{
At the glass transition the correlation length for parallel
orientation obtained from the half-width of the peak of $S_{22}(q,0)$
is plotted as a function of the aspect ratio. For X$_0 \leq 2.4$ 
the glass transition is of type B whereas for X$_0 \geq 2.5$ it is of
type B' (see text).
}}
\label{fig:4}
\end{figure}

\begin{figure}
\unitlength1cm
\epsfxsize=10.5cm
%\begin{picture}(7,7.5)
%\put(-2.2,-0.5){\rotate[r]{\epsffile{fig3.ps}}}
%\end{picture}
\caption{{
For $X_0=2.5$ and $\phi=0.593$, the smallest aspect ratio where the
type-B' transition occurs first the q-dependence of the non-ergodicity
parameter $F'_{ll'}(q,m)$ is plotted 
together with the static structure factor $S_{ll'}(q,m)$ and the
normalized non-ergodicity parameter $f'_{ll'}(q,m) = F'_{ll'}(q,m) /
S_{ll'}(q,m)$. In (a) this is done for $l=l'=2$ and $m=0$ whereas (b)
shows the same quantities for the center of mass $l=l'=m=0$.  
}}
\label{fig:6}
\end{figure}

\begin{figure}
\unitlength1cm
\epsfxsize=10.5cm
%\begin{picture}(7,7.5)
%\put(-2.2,-0.5){\rotate[r]{\epsffile{fig3.ps}}}
%\end{picture}
\caption{{
The formation of the $\phi_c^{(B')}$ glass transition is illustrated. 
Within each domain
of diameter $\xi$ the system shows liquid crystalline order whereas
for $l \gg \xi$ there are randomly frozen orientational correlations. 
}}
\label{fig:5}
\end{figure}

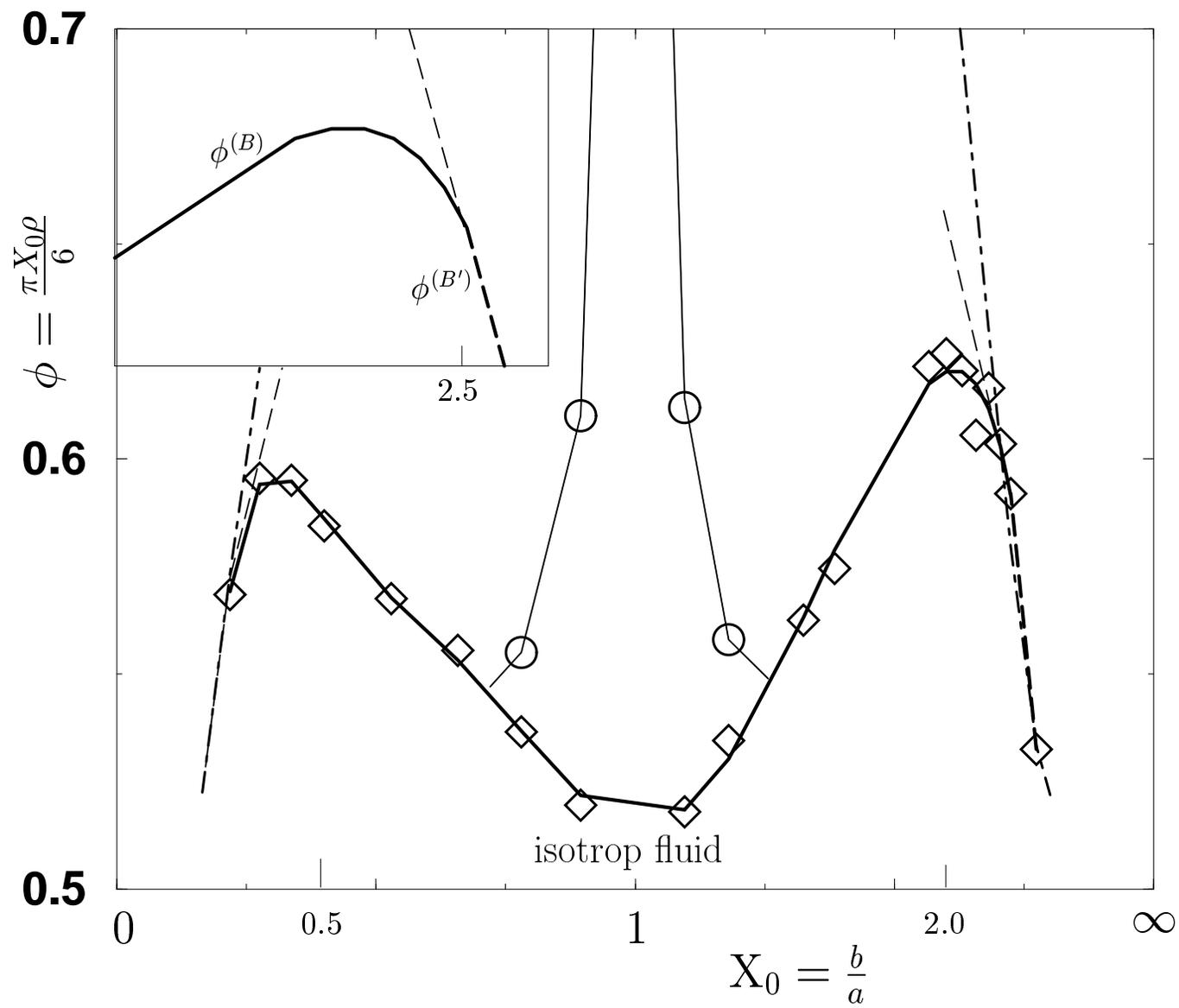
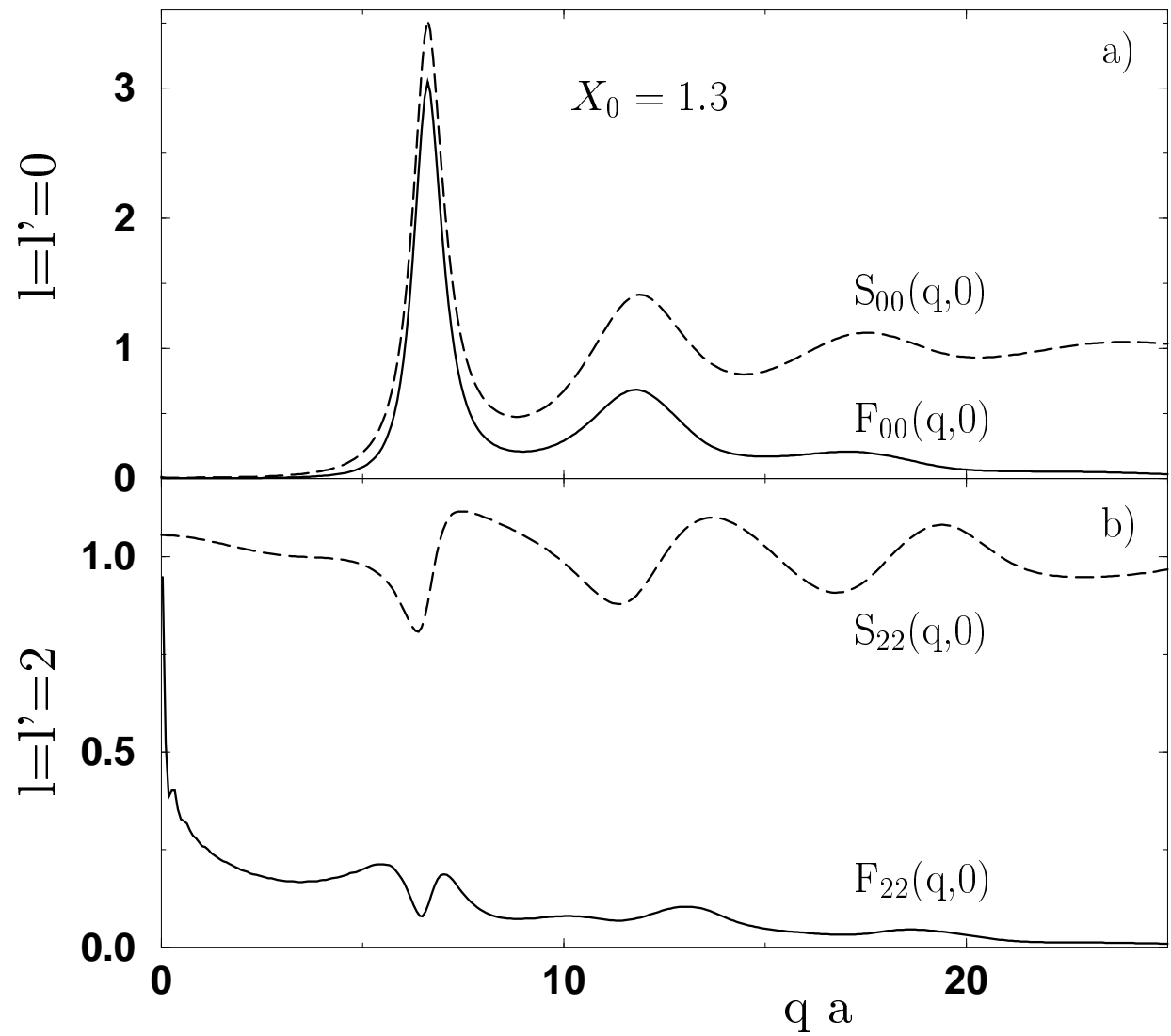
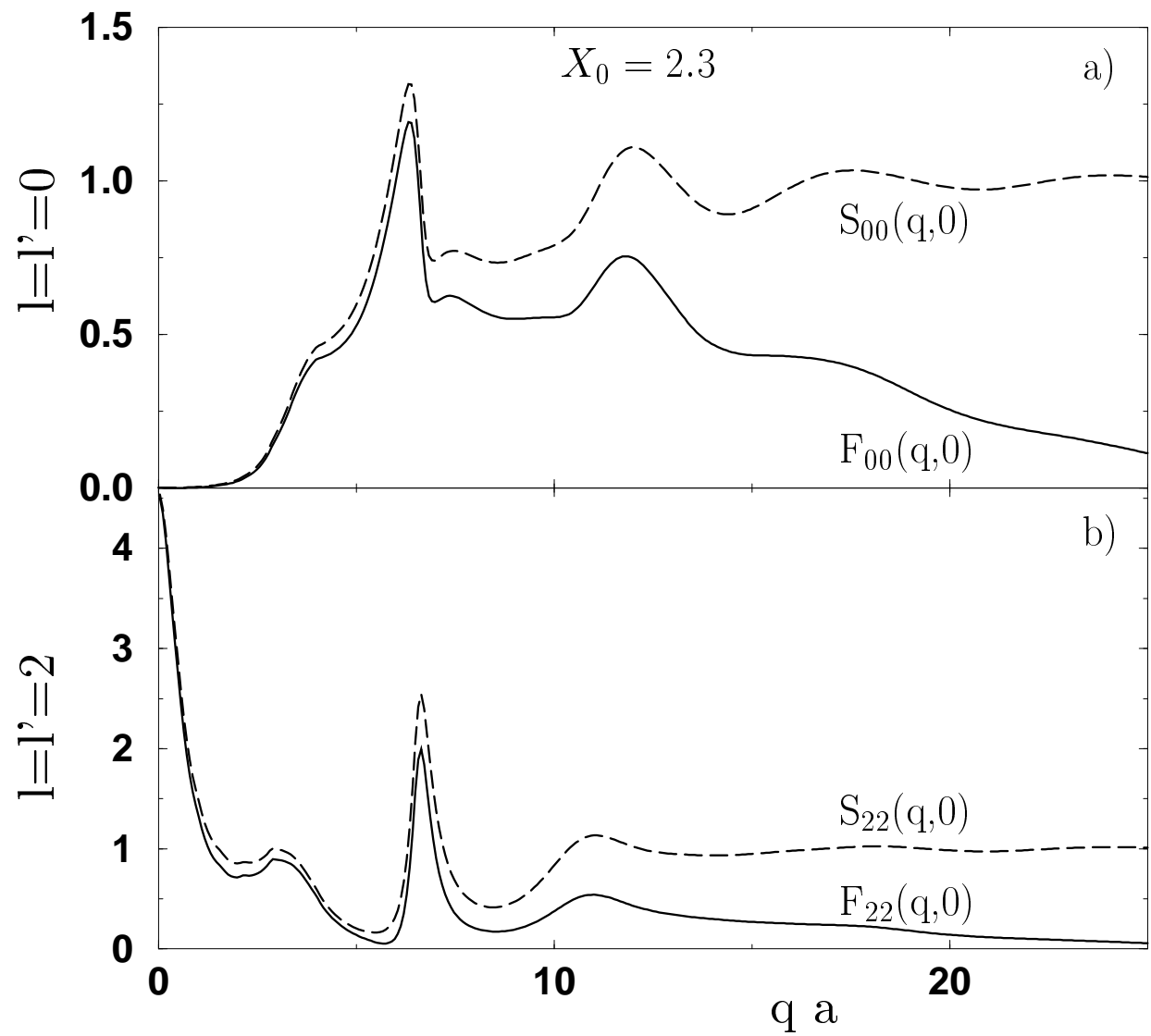
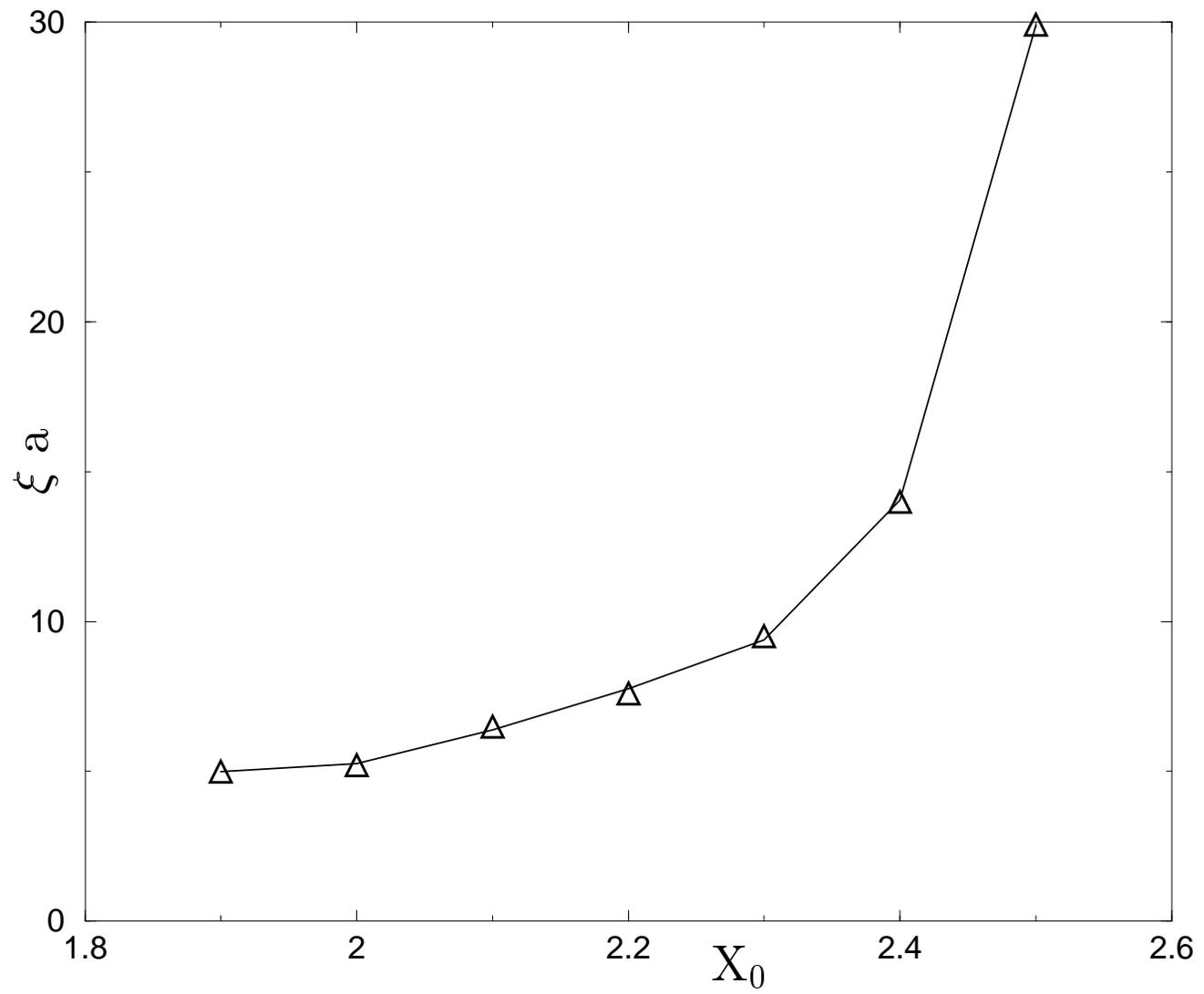
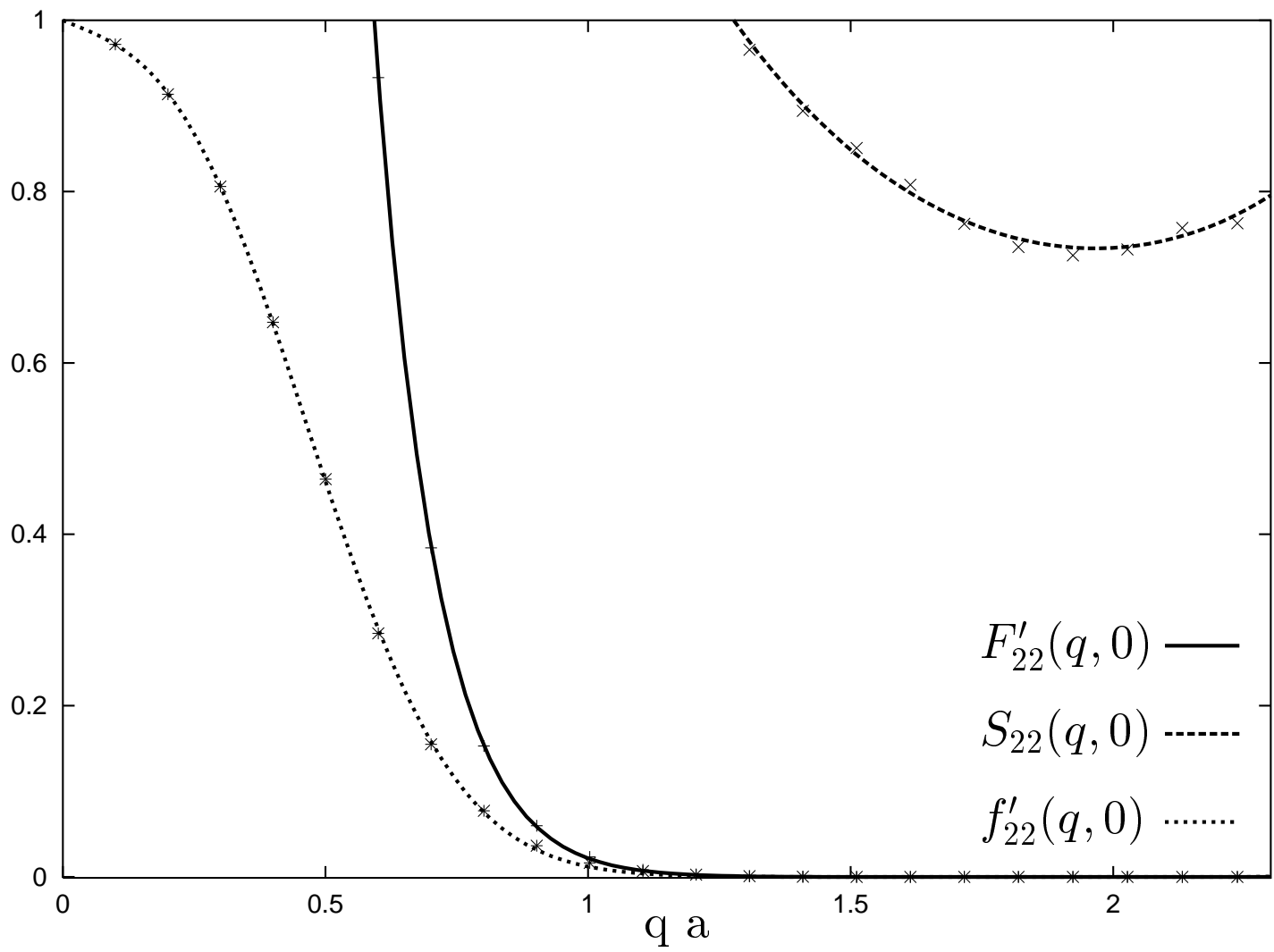
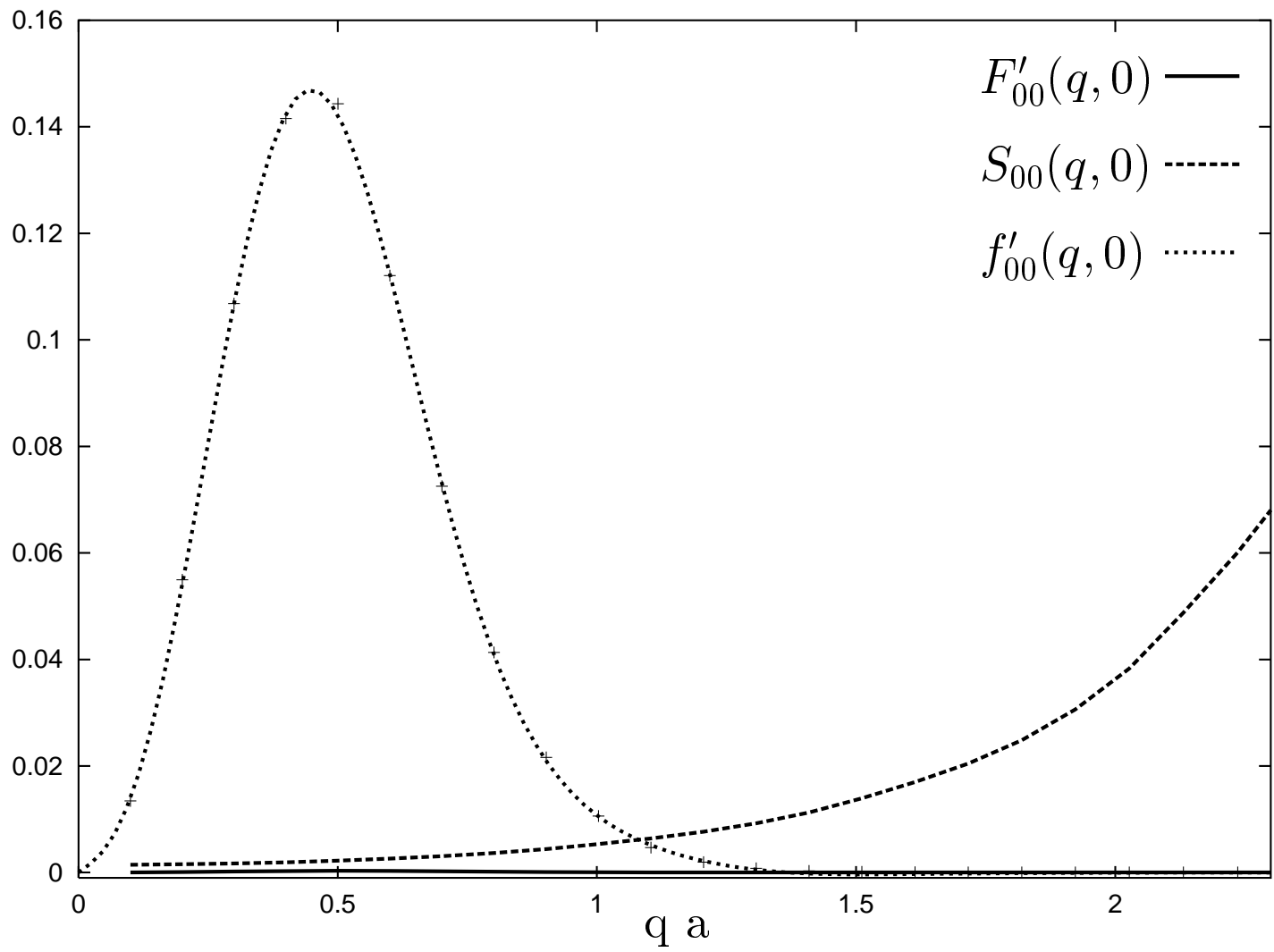
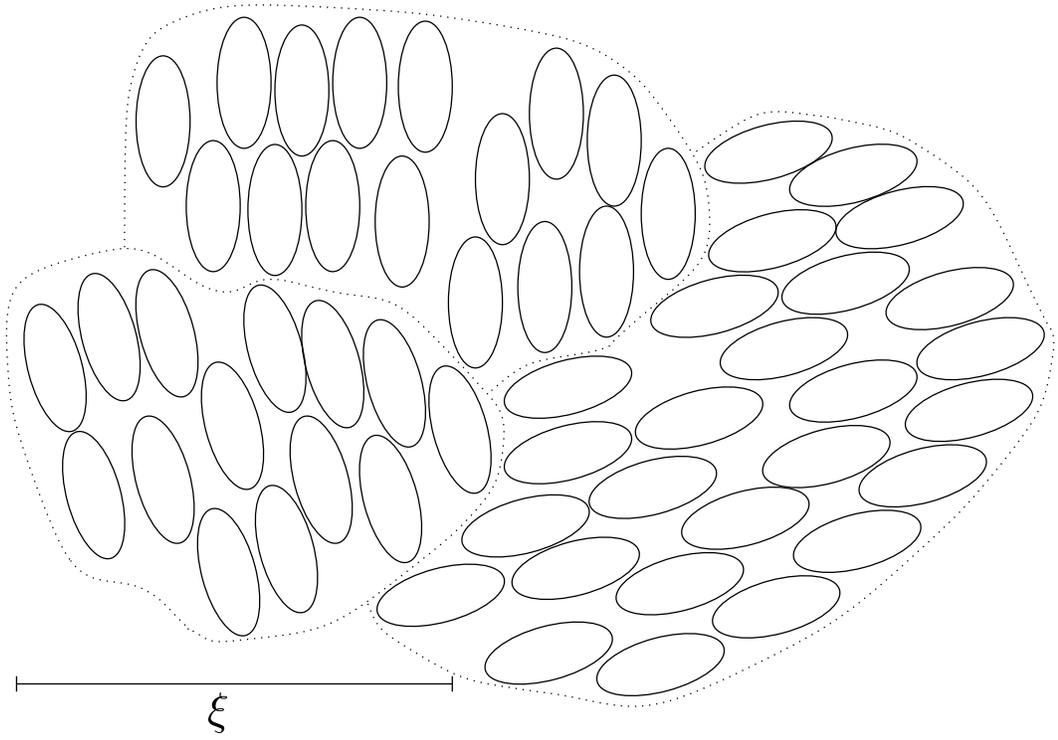
\end{document}